\documentclass[preprint2,numberedappendix]{emulateapj-rtx4}
\usepackage{natbib}
\usepackage{amsmath}
\usepackage{graphicx,times,bm,url}
\usepackage{afterpage,lscape}
\graphicspath{{./fig/}{./png/}}
\newcommand{\EQ}{\begin{equation}}
\newcommand{\EN}{\end{equation}}
\newcommand{\EQA}{\begin{eqnarray}}
\newcommand{\ENA}{\end{eqnarray}}

\newcommand{\Fig}[1]{Figure~\ref{#1}}

\newcommand{\Tab}[1]{Table~\ref{#1}}

{}
{}
{}

{}
{}
{}
{}
{}
{}
{}
{}
{}
{}
{}
{}
{}
{}
{}
{}
{}
{}
{}
{}
{}

{}
{}
{}

{}

{}
{}

\newcommand{\eee}{\hat{\mbox{\boldmath $e$}} {}}

\newcommand{\gggg}{\mbox{\boldmath $g$} {}}

\newcommand{\rr}{\mbox{\boldmath $r$} {}}

\newcommand{\BB}{\bm{B}}
\newcommand{\HH}{\bm{H}}

\newcommand{\uu}{\mbox{\boldmath $u$} {}}

\newcommand{\JJ}{\mbox{\boldmath $J$} {}}

\newcommand{\AAA}{\mbox{\boldmath $A$} {}}

\newcommand{\ee}{\mbox{\boldmath $e$} {}}
\newcommand{\nn}{\mbox{\boldmath $n$} {}}

\newcommand{\FF}{\mbox{\boldmath $F$} {}}

\newcommand{\nab}{\mbox{\boldmath $\nabla$} {}}
\newcommand{\SSSS}{\mbox{\boldmath ${\sf S}$} {}}

\newcommand{\DD}{{\rm D} {}}

\newcommand{\dd}{{\rm d} {}}

\def\Rm{\mbox{\rm Re}_{\rm M}}

\def\Rey{\mbox{\rm Re}}

\def\cp{c_{p}}
\def\cv{c_{v}}

\def\cs{c_{\rm s}}

\def\Hm{H_{\rm m}}

\newcommand{\cm}{\,{\rm cm}}

\usepackage{color}

\usepackage{siunitx}

\DeclareSIUnit\parsec{pc}
\DeclareSIUnit\lightyear{ly}
\DeclareSIUnit\year{yr}
\DeclareSIUnit\gauss{G}

\begin{document}

\title{Stabilizing effect of magnetic helicity on magnetic cavities in the intergalactic medium}

\author{Simon Candelaresi}
\affiliation{School of Mathematics and Statistics, University of Glasgow, Glasgow G12 8QQ, United Kingdom \\ simon.candelaresi@gmail.com}
\affiliation{Division of Mathematics, University of Dundee, Dundee DD1 4HN, United Kingdom}

\author{Fabio Del Sordo}
\affiliation{Institute of Astrophysics, FORTH, GR-71110 Heraklion, Greece}
\affiliation{Department of Physics, University of Crete, GR-70013 Heraklion, Greece\\ fabiods@ia.forth.gr}

\begin{abstract}
We investigate the effect of magnetic helicity on the stability of buoyant magnetic cavities
as found in the intergalactic medium.
In these cavities we insert helical magnetic fields and test whether or not helicity
can increase their stability to shredding through the Kelvin-Helmholtz instability
and, with that, their life time.
This is compared to the case of an external vertical magnetic field which is known to reduce the
growth rate of the Kelvin-Helmholtz instability.
By comparing a low-helicity configuration with a high helicity one with the same magnetic energy
we find that an internal helical magnetic field stabilizes the cavity.
This effect increases as we increase the helicity content.
Stabilizing the cavity with an external magnetic field requires instead a significantly
stronger field at higher magnetic energy.
We conclude that the presence of helical magnetic fields
is a viable mechanism to explain
the stability of intergalactic cavities on time scales longer than $\SI{100}{\mega\year}$.
\end{abstract}

\section{Introduction}

Intergalactic hot cavities have been observed to emanate from galactic disks in clusters of galaxies
(e.g.\ \cite{Carilli-Perley-1994-270-173-MNRAS, Carilli-Taylor-2002-40-319-ARAA, Taylor-Fabian-2002-334-769-MNRAS,
Churazov-Bruggen-2001-554-261-ApJ, Birzan2004ApJ, McNamara-Nulsen-2007-45-117-ARAA, Montmerle-2011-51-299-EAS}).
Similar but different structures have been observed also around our Galaxy, they are named Fermi Bubbles
\citep{Su2010ApJ} and their formation mechanism is still elusive \citep{Yang-Ruszkowski-2018-6-1-MDPI}.
Observations seem to indicate that such intergalactic structures form as consequences of radio jets emanated
by supermassive black holes or by active galactic nuclei (AGN) and their interaction with extragalactic plasma.
These AGN-inflated radio bubbles in the intergalagtic medium are seen in X-ray images of galaxy clusters.
A shock is produced as the jet penetrates the surrounding medium, and, in order to achieve pressure
equilibrium, the jet material expands leading to the formation of a low-density cavity.
Bubbles at \SI{10}{\kilo\parsec} from the galactic center are found to be at least a factor of $3$
less dense than the surrounding medium.

These hot cavities propagate through the intergalactic medium where they are subject
to magnetohydrodynamical instabilities, such as the Rayleigh-–Taylor instability, the Richtmyer-–Meshkov instability,
and in particular, the Kelvin--Helmholtz (KH) instability, after which turbulent mixing occurs.
However, estimates of their life time using their terminal velocity (e.g.\ \cite{Birzan2004ApJ}) suggests
that they survive significantly longer than they should (of the order of $\SI{10}{\mega\year}$-$\SI{100}{\mega\year}$).

The intergalactic medium can be modeled as high-conductivity plasma.
\cite{Chandrasekhar1961hhs, Sharma-Srivastava-1968-21-917-AusJPhys} showed analytically
how in the presence of a magnetic field parallel to the velocity the KH instability is suppressed.
However, a perpendicular field has no effect on the growth of the modes.
The effect of an external magnetic field on the stability of the cavities has been subject of several studies
\citep[e.g.][]{Robinson-Dursi-2004-601-621-ApJ}.
However, the interiors of these cavities may be magnetized too.
This is possible, for instance, if the jet responsible for their inflation is magnetized.
Also, there is the possibility that the turbulent motions arising during the generation of
the cavities may amplify the magnetic field in its interior through e.g.\ a dynamo effect.
The hypothesis of a helical magnetic field lying inside these bubbles appears justified also in light of results
showing that AGN jets are characterized by helical magnetic fields \citep{Li2006ApJ, Tang2008ApJ, Gabuzda2018}.
If this field contains magnetic helicity, it is stable on diffusive times, i.e.\ the
energy and topology change insignificantly during this period \citep[e.g.,][]{DS2010PhRvE, C2011PhRvE}.

If such an internal field was not stretched throughout the cavity boundaries where the KH instability
occurs, it would have no immediate effect on the instability and we would not see a direct suppression of it.
However, as the instability grows and leaves the linear regime, it can potentially be anyway suppressed
at this later stage.

Intergalactic cavities have been simulated in the past and different effects were considered in
order to understand their evolution and stability.
For instance, \cite{Bruggen-2003-592-839-ApJ} studied the cooling behavior in buoyant bubbles
in galactic clusters, where there is a significant growth of the KH instability.
The inflation process of the bubble also plays a role in determining the bubble's stability.
\cite{PS2006MNRAS} illustrated how the initial deceleration and drag, albeit unable to prevent the
disruption of a bubble, may significantly lengthen a cavity's lifetime, and \cite{SS2008MNRAS} proposed
such dynamics as the explanation for long-lasting cavities to exist.
\cite{Braithwaite2010MNRAS} studied the magnetohydrodynamic relaxation of AGN ejecta, finding that the time
scale on which the bubble reaches an equilibrium depends on the magnetization and the helicity of the outflow.
It should be noted however, that this study did not include the effect of buoyancy.
Subsequently, \cite{Gourgouliatos2010MNRAS} studied the problem with both analytical and numerical approaches,
finding that the presence of both poloidal and toroidal components of the magnetic field in intergalagtic cavities
does indeed induce stability.
\cite{Liu-Hsu-2008-15-7-PhysPlas} showed how the ratio of the toroidal to poloidal magnetic field of the bubble
determines the direction of bubble expansion and propagation, which can develop asymmetries about its propagation axis.
\cite{Dong-Stone-2009-704-1309-ApJ} studied the effect of magnetic fields and anisotropic viscosity,
finding that a critical role for the evolution of the bubble is played by the initial field geometry,
and that toroidal field loops initially located inside the bubble are the best option to
reproduce the observed cavity structures.

\cite{Vogt2005AA} observed how the typical length scale of magnetic fields in the
Hydra A cluster is smaller than the typical bubble size, therefore concluding that the
scenario of an uniform external field is not supported.
Based on this result, \cite{Ruszkowski-Ensslin-2007-378-662-MNRAS} studied the effect of a
random magnetic field, taking into account both the helical and non-helical cases.
They argued that when the gas pressure is higher than the magnetic pressure, that is for
high plasma $\beta$ ($\sim 40$ in their case), a random helical magnetic field cannot
stabilize the bubble.
Still, they could not exclude that stabilization is taking place if plasma $\beta$ is
locally lower inside the intergalactic cavities.
On the other hand, \cite{Jones-DeYoung-2005-624-586-ApJ} found with 2D simulations that
micro Gauss magnetic fields can stabilize these bubbles.
In general, the evolution of the helicity of extragalactic bubbles and its
interaction with the intergalactic medium, after the bubble's inflation and detachement from the jet,
have been only marginally studied and they are not yet well understood,  especially in three dimensions.
Therefore, further investigation is needed, and the research we here present intends to be a step in this direction.

The aim of this work is to investigate the stabilizing effects of internal helical magnetic fields
on the intergalactic cavities where they may be harbored. 
Moreover, we aim at comparing their stabilizing effect to that of a homogeneous
external vertical field to evaluate whether a similar stabilizing effect may be attained with weaker fields.
This work is organized as follows:
In Section \ref{sec: model} we illustrate the model we implemented; 
In Section \ref{sec: internal} we discuss the effect of an internal helical magnetic field
for two different kinds of initial magnetic field;
In Section \ref{sec: external} we show the effect of an external magnetic field;
and in Section \ref{sec: conclusion} we draw some conclusions and implications of this work.

\section{Model}
\label{sec: model}

Our model set-up consists of a hot under-dense bubble 
%located 
embedded in a stably stratified
medium with gravity.
This bubble rises through buoyancy, which leads to shear with the surrounding stationary medium
and the onset of the Kelvin--Helmholtz instability on the bubble's surface.
Our goal is to study how the stability of the bubble depends on the presence of magnetic fields.
Therefore, concerning the stability of the bubbles we will consider four cases:
(i) purely hydrodynamical case, (ii) internal ABC helical magnetic field, 
(iii) internal spheromak magnetic field, and (iv) external vertical magnetic field.
The last scenario has already been studied for the Kelvin--Helmholtz instability
for incompressible and compressible media by \cite{Chandrasekhar1961hhs} and
here we will use it as reference.
Cases (ii) and (iii) will be further subdivided into a high-helicity and low-helicity case, while for case (iv) we will take into account
strong and weak external field cases.

\subsection{Governing Equations}

To approach this problem and to provide a quantitative assessment we make use of direct numerical simulations, that we
perform with the public code {\sc PencilCode}
(\url{https://github.com/pencil-code}).
This code is particularly suitable for our study since it avoids using the magnetic field
$\BB$ as primary variable and instead uses its vector potential $\AAA$, hence ensuring the fields
stay solenoidal throughout the simulations.
Moreover it allows to quantify
the magnetic helicity $H=\int\AAA\cdot\BB\ \dd V$, where the integral is calculated over the whole computational domain.
For our study we require an ideally conducting viscous medium.
This is governed by the resistive magnetohydrodynamics equation together with the energy
(temperature) equation:

\EQA
\frac{\partial \AAA}{\partial t} & = & \uu\times\BB + \eta\nab^2\AAA,
\label{eq: induction} \\
\frac{\DD \uu}{\DD t} & = & -\cs^{2}\nab \left( \frac{\ln T}{\gamma} + \ln{\rho} \right)
 + \frac{\JJ\times\BB}{\rho} \nonumber \\
 & &  -\gggg + \FF_{\rm visc},
\label{eq: momentum} \\
\frac{\DD \ln{\rho}}{\DD t} & = & -\nab \cdot \uu,
\label{eq: continuity} \\
\frac{\partial \ln T}{\partial t} & = & -\uu\cdot\nab\ln T - (\gamma - 1)\nab\cdot\uu \nonumber \\
 & & + \frac{1}{\rho c_v T} \left( \nab\cdot(K\nab T) + \eta\JJ^2 \right. \nonumber \\
 & & \left. + 2\rho\nu\SSSS\otimes\SSSS + \zeta\rho(\nab\cdot\uu)^2 \right),
\label{eq: temperature}
\ENA
with the magnetic vector potential $\AAA$,
magnetic field $\BB = \nab\times\AAA$,
fluid velocity $\uu$,
constant magnetic resistivity (diffusivity) $\eta$,
advective derivative $\DD / \DD t = \partial/\partial t + \uu\cdot\nab$,
sound speed $\cs = \gamma p/\rho$,
adiabatic index $\gamma = \cp/\cv$,
heat capacities $\cp$ and $\cv$ at constant pressure and volume,
temperature $T$,
density $\rho$,
electric current density $\JJ = \nab\times\BB$,
gravitational acceleration $\gggg$,
viscous force $\FF_{\rm visc}$,
heat conductivity $K$
and the bulk viscosity $\zeta$.
The viscous force is given as $\FF_{\rm visc} = \rho^{-1}\nab\cdot 2\nu\rho\SSSS$,
with the traceless rate of strain tensor $S_{ij} = \frac{1}{2}(u_{i,j} + u_{j,i}) - \frac{1}{3}\delta_{ij}\nab\cdot\uu$.
The equation of state used here is for the ideal monatomic gas and it appears
implicitly in our equations, as we eliminated pressure $p$.
Here the gas is monatomic with $\gamma = 5/3$.

Our side boundaries ($xy$) are chosen to be periodic, while the bottom is closed ($\uu\cdot\nn = 0$)
and the top open.
This allows for fluxes outwards.
For the magnetic field the vertical boundaries are set to open, allowing for magnetic flux.

As simulation domain we choose a box of size $2.4\times2.4$ in the horizontal ($xy$) plane
and $9.6$ in the vertical ($z$), using $480\times480\times1920$ meshpoints.

In order to reduce resistive magnetic helicity decay we choose a value of the magnetic resistivity
as low as the resolution allows.
Here we set it to $\eta = 3\times 10^{-4}$.
Viscosity is either set to $\nu = 1\times 10^{-3}$ 
or to $\nu = 2\times 10^{-4}$
to check that the overall behavior of the simulation does not
depend on this parameter
which prevents any accumulation of turbulent
energy at small scales.

With the viscosity and magnetic resistivity we can then compute our Reynolds numbers:
\EQA
\label{Eq:Rey}
\Rey & = & \frac{u_{\rm max}d}{\nu} \label{eq: Re} \\
\Rm & = & \frac{u_{\rm max}d}{\eta} \label{eq: Rm} ,
\ENA
with the maximum velocity $u_{\rm max}$ and bubble diameter $d = 1.6$.

\subsection{Initial Thermodynamic Conditions}

As initial condition we choose a stably stratified atmosphere in which we
place an under-dense hot cavity of spherical shape.

This can be found as part of the {\sc PencilCode} under
\path{src/initial_condition/bubbles_init.f90}.
The stable atmosphere obeys the hydrostatic equilibrium
\EQ\label{eq: hydrostatic equilibrium}
\rho g = -\nab p,
\EN
with the gravitational acceleration in negative $z$-direction $g$.
Our model atmosphere extends in the $z$-direction which makes the gradient
a derivative in $z$ and we can rewrite equation \eqref{eq: hydrostatic equilibrium} as
\EQ \label{eq: hydrostatic equilibrium z}
\rho g = -\frac{\dd p}{\dd z}.
\EN
Density, pressure and temperature are related through the ideal gas law
\EQ\label{eq: ideal gas}
p = \frac{R}{\mu} \rho T,
\EN
where $\mu$ is the mass of one mol of gas and
$R = \SI{8.31}{\joule\per\kelvin\per\mole}$
is the ideal gas constant.
With the ideal gas law we can express the hydrostatic equilibrium as
\EQ\label{eq: hydrostatic equilibrium T p}
\frac{\mu g}{R}\frac{\dd z}{T} = -\frac{\dd p}{p}
\EN

The gas is chosen to be adiabatic, i.e.\
\EQ\label{eq: adiabatic}
p^{1-\gamma}T^{\gamma} = {\rm const.},
\EN
which leads to the relation
\EQ\label{eq: dp dT}
\frac{\dd p}{p} = \frac{\gamma}{\gamma-1}\frac{\dd T}{T}.
\EN
With equation \eqref{eq: hydrostatic equilibrium T p} we can write
\EQ
\frac{\dd T}{\dd z} = -\frac{\gamma-1}{\gamma}\frac{\mu g}{R}.
\EN
We now integrate this equation and obtain
\EQ
T = -\frac{\gamma-1}{\gamma}\frac{\mu g}{R} z + T_0,
\EN
where $T_0$ is the temperature at an arbitrary height $z_0$.

We now choose this height to be the isothermal scale height
\EQ\label{eq: isothermal scale height}
z_0 = \frac{RT_0}{\mu g}.
\EN
This is justified if we assume that the gas is in isothermal equilibrium at $z_0$,
which is a common assumption for the adiabatic atmosphere.
Our temperature profile obtains now the form:
\EQ\label{eq: profile T}
T(z) = T_{0}\left( 1-\frac{\gamma-1}{\gamma} \frac{z}{z_{0}} \right).
\EN

With the temperature profile and equation \eqref{eq: dp dT} we can compute the
pressure profile to
\EQ\label{eq: profile p}
p(z) = p_0\left( 1-\frac{\gamma-1}{\gamma} \frac{z}{z_{0}} \right)^{\gamma/(\gamma-1)}.
\EN
Taking equation \eqref{eq: hydrostatic equilibrium z} we can also compute the density profile to
\EQ\label{eq: profile rho}
\rho(z) = \rho_0\left( 1-\frac{\gamma-1}{\gamma} \frac{z}{z_{0}} \right)^{1/(\gamma-1)}.
\EN

For the cavity we choose a radius of $0.8$ and place it centrally in $x$ and $y$ and at $0.8$ in $z$.
Its initial peak temperature is $T_{\rm cavity} = 4$ and density $\rho_{\rm cavity} = 0.25$.
In order to avoid sudden wave formation we apply (sharp) smoothing for $T_{\rm cavity}$ and
$\rho_{\rm cavity}$ near the edges using a $\tanh$ profile that match the surrounding values.
With that the temperature and density is constant up to ca.\ $72\%$ of the radius of the cavity.

This should be contrasted to the surrounding medium which is stably stratified with
gravitational acceleration of $g = 0.1$.
For that we choose $z_0 = 4$, $\rho_0 = 1$ and $T_0 = 1$ such that the cavity is in
approximate pressure balance with its surrounding medium and its expansion
or compression is insignificant.

\subsection{Initial Magnetic Condition 1: ABC Field}
\label{Sec:ABC}
 For the simulations that include a magnetic field inside the bubble we consider two kinds of initial magnetic conditions.
In the first case we insert
a helical magnetic field of the Arnold-Beltrami-Childress (ABC) flow type:
\EQ
\AAA = f(r) A_0 \left(
\begin{array}{c}
\cos((y-y_{\rm cavity})k) + \sin((z-z_{\rm cavity})k) \\
\cos((z-z_{\rm cavity})k) + \sin((x-x_{\rm cavity})k) \\
\cos((x-x_{\rm cavity})k) + \sin((y-y_{\rm cavity})k)
\end{array}
\right),
\label{eq: abc}
\EN
with the function
\EQ
f(r) = 1 - (r/r_{\rm b})^{n_{\rm smooth}}
\EN
that makes sure that the magnetic field vanishes outside the bubble without any strong current sheets.
Here $r = \sqrt{(x-x_{\rm cavity})^2 + (y-y_{\rm cavity})^2 + (z-z_{\rm cavity})^2}$ and
$r_{\rm b}$ is the radius of the bubble.
We use as smoothing coefficient $n_{\rm smooth} = 2$.
For values of $r > r_{\rm b}$ we set the field to $0$.

Taking the curl we obtain $\BB = k\AAA$.
The magnetic energy scales like $A_0^2 k^2$, while the magnetic helicity scales like $A_0^2 k$.
By inversely scaling the amplitude $A_0$ and the inverse scale $k$ we can change the
magnetic helicity while keeping the magnetic energy fixed.
For our low-helicity set up we choose $A_0 = 2.5\times10^{-2}$ and $k = 20$.
For the high helicity case we choose $A_0 = 0.1$ and $k = 5$.
For simplicity we will occasionally write $H = 1$ and $H = 4$ for the two cases.

\subsection{Initial Magnetic Condition 2: Spheromak Field}
\label{Sec:Spheromak}
To make sure our results are not dependent on a specific initial geometry of the internal magnetic field,
but only the topology described by the amount of magnetic helicity,
we perform a second set of simulations that make use of a spheromak magnetic field.
For that we use the form given by \cite{Aly-Amari-2012-420-237-MNRAS}.
There the magnetic field is given in spherical coordinates $(r, \theta, \phi)$,
with origin at the bubble's center, as
\EQA
\BB & = & 2A_0\frac{g(\alpha r)}{(\alpha r)^2}\cos{(\theta)}\eee_r \nonumber \\
 & & - A_0\frac{g'(\alpha r)}{\alpha r}\sin{(\theta)}\eee_\theta \nonumber \\
 & & + A_0 \frac{g(\alpha r)}{\alpha r}\sin{(\theta)}\eee_\phi, \label{eq: bb spheromak}
\ENA
with $\alpha = \tau/r_{\rm b}$.
The function $g$ is given as
\EQ
g(t) = \frac{t^2}{\tau^2} - \frac{3}{\tau\sin{(t)}} \left( \frac{\sin{(t)}}{t} - \cos{(t)} \right)
\EN
and its derivative as
\EQ
g'(t) = \frac{2t}{\tau^2} - \frac{3}{\tau\sin{(t)}} \left( -\frac{\sin{(t)}}{t^2} + \frac{\cos{(t)}}{t} + \sin{(t)} \right),
\EN
with $\tau$ being the zeros of the equation $g(\tau) = 0$ that can be calculated numerically to
$\tau_1 = 5.763$, $\tau_2 = 9.095$, $\tau_3 = 12.229$, $\tau_4 = 15.515$ and $\tau_5 = 18.689$
for the first five roots.
We then perform the transformation into Cartesian coordinates using the relations
\EQA
B_x & = & B_r(x-x_{\rm cavity})/r \nonumber \\
 & & + B_{\theta}(x-x_{\rm cavity})(z-z_{\rm cavity})/(r_{xy}r) \nonumber \\
 & & - B_{\phi}(y-y_{\rm cavity})/r_{xy} \\
B_y & = & B_r(y-y_{\rm cavity})/r_{\rm b} \nonumber \\
 & & + B_\theta(y-y_{\rm cavity})(z-z_{\rm cavity})/(r_{xy}r) \nonumber \\
 & & + B_\phi(x-x_{\rm cavity})/r_{xy} \\
B_z & = & B_r(z-z_{\rm cavity})/r_{\rm b} - B_\theta r_{xy}/r
\ENA
with $r_{xy} = \sqrt{(x-x_{\rm cavity})^2 + (y-y_{\rm cavity})^2}$.

\cite{Aly-Amari-2012-420-237-MNRAS} derived this field for a cylindrically symmetric
force balance between magnetic and pressure forces solving the Grad--Shafranov equation.
However, we notice that in our case the hydrostatic pressure is not constant in space.
Nevertheless, since the pressure deviation from $p_0$ is of the order of $0.5\%$ we assume it to be constant in our bubble.

\begin{figure}[t!]\begin{center}
\includegraphics[width=0.95\columnwidth]{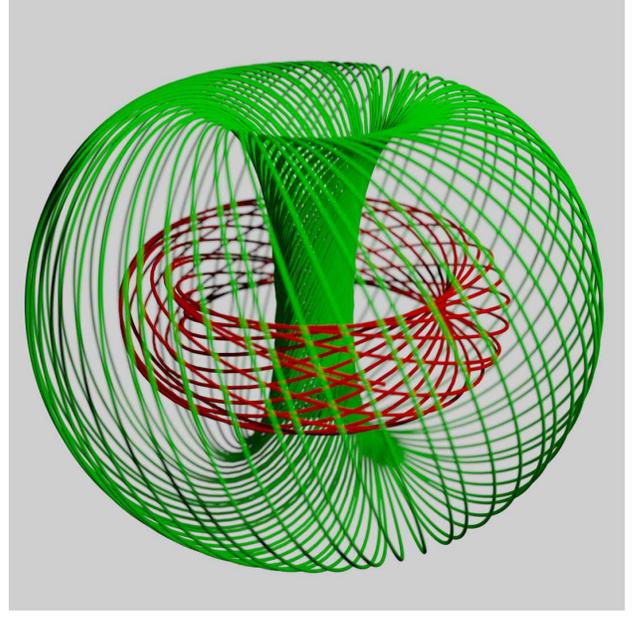}
\end{center}
\caption[]{
Rendering of two magnetic field lines for the spheromak initial condition.
}
\label{fig: spheromak}
\end{figure}

We restrict the field within the bubble by setting it to $0$ outside.
We can do this, as the field naturally and smoothly vanishes at $r = r_{\rm b}$
without generating any strong surface currents.
To obtain the magnetic vector potential to this field we solve the equation
$\nab^2\AAA = -\nab\times\BB$ in Fourier space, where we use the Coulomb gauge ($\nab\cdot\AAA = 0$)
and the fact that the field is periodic, since it vanishes at the boundaries.
A rendering of the magnetic field can be seen in \Fig{fig: spheromak}.

\subsection{Initial Magnetic Condition 3: Vertical Field}
\label{Sec:Vertical}

The third magnetic case we take into consideration is that of an initial vertical magnetic field
\EQ
\BB = B_0 \ee_z
\EN
where $B_0$ is the initial amplitude of the field
in the entire computational domain.

\subsection{Test Cases}

For a fair comparison, and a quantitative evaluation of the role of
magnetic fields in the three cases described in Sections \ref{Sec:ABC}, \ref{Sec:Spheromak}, and \ref{Sec:Vertical},
we first perform calculations of a purely hydrodynamical case with no magnetic fields.
We compare this with the two cases with an internal magnetic field within the cavities.
The two ABC cases have almost the same magnetic energy of $E_{\rm m} = 0.2182$,
while the two spheromak cases have energies of $E_{\rm m} = 0.264$ for the high helicity
and $E_{\rm m} = 0.189$ for the low helicity case.
This difference is a result from the adjustments of the parameters that are chosen
such to match the magnetic helicity content of the ABC case as closely as possible,
while keeping the difference with the magnetic energy minimal.
Here the high helicity case as a four times larger magnetic helicity content.
An overview over the parameters used in the simulations is show in \Tab{tab: simulations}.
With these initial conditions we can compute the plasma $\beta$ where the magnetic
field is strongest using
\EQ\label{eq: plasma beta}
\beta = \min\left( \frac{2(R \rho T / \mu)}{B^2} \right).
\EN
We compute this to be $\beta \approx 0.6$ for the helical ABC cases
and $\beta = 0.44$ for the high helicity spheromak and $\beta = 0.038$ for the
low helicity spheromak case.
However, we have to notice that this is the minimum value $\beta$ attains in our model since,
with the presence of magnetic null points within the domain, $\beta$ is not constant and even diverges at some points.
In the external field cases we have instead $\beta = 20$ (weak field) and $\beta = 1.25$ (strong field).
% magnetic beta (minimum) np.min(np.exp(var.lnTT)*var.rho/pc.math.dot2(var.bb)*2*0.4), with R = 0.4.
% rise_helical_nu_2e-4e_sharp_highRes, beta = 0.6337188708922615, Hm = 0.07353123914253609, Em = 0.263963545145017
% rise_helical_nu_2e-4f_sharp_highRes, beta = 0.5410630476096924, Hm = 0.01838280763483448, Em = 0.18883641451418123
% rise_helical_nu_2e-4e_spheromak, beta = 0.4421477, Hm = 0.07186382, Em = 0.2181704193353653
% rise_helical_nu_2e-4f_spheromak, beta = 0.03840602, Hm = 0.018923093, Em = 0.2181699424982071

\begin{table}
\centering
\begin{tabular}{llllllll}
Model & $\BB (A_0)$ & $\Hm$ & $\nu$ & $\eta$ & $k$/$\tau$ & Re & Re$_{\rm M}$ \\ \hline 
hydro & - & - & $10^{-3}$ & - & - & $960$ & - \\
hydro2 & - & - & $2\times10^{-4}$ & - & - & $4800$ & - \\
hel\_l & 0.025 & 1 & $10^{-3}$ & $3\times 10^{-4}$ & $20$ & $1280$ & $4200$ \\
hel\_h & 0.1 & 4 &$10^{-3}$ & $3\times 10^{-4}$ & $5$ & $1280$ & $4200$ \\
hel\_l2 & 0.025 & 1 & $2\times10^{-4}$ & $3\times 10^{-4}$ & $20$ & $5600$ & $3700$ \\
hel\_h2 & 0.1 & 4 &$2\times10^{-4}$ & $3\times 10^{-4}$ & $5$ & $6400$ & $4200$ \\
sph\_l & 6.39 & 1 & $2\times10^{-4}$ & $3\times 10^{-4}$ & $21.85$ & $7200$ & $4800$ \\
sph\_h & 1.7 & 4 &$2\times10^{-4}$ & $3\times 10^{-4}$ & $5.76$ & $11000$ & $7500$ \\
ex\_low & 0.2 & 0 & $10^{-3}$ & $3\times 10^{-4}$ & - & $320$ & $1000$ \\
ex\_high & 0.8 & 0 & $10^{-3}$ & $3\times 10^{-4}$ & - & $320$ & $1000$ \\
\hline 
\end{tabular}
\caption{
List of simulations with parameters used in this study: The magnetic field intensity, measured by the parameter $A_0$ in equation \eqref{eq: abc},
the magnetic helicity $\Hm$, the viscosity $\nu$, the magnetic diffusivity $\eta$, the parameter $k$ of the ABC flow in
equation \eqref{eq: abc} (parameter $\tau$ of the spheromak configuration in equation \eqref{eq: bb spheromak}),
the Reynolds number $\Rey$ and the magnetic Reynolds number $\Rm$, as defined in equations \eqref{eq: Re} and \eqref{eq: Rm}.
}
\label{tab: simulations}
\end{table}

\subsection{Unit Conversions}

We choose the conversion rate from code units to physical units
such that the dimensions of the setups correspond to physically
observed numbers in the intergalactic medium (see \Tab{tab: unit conversions}).

With these code unit conversions our set up has a size of $L_{xy} = \SI{24}{\kilo\parsec}$ horizontally
and $L_z = \SI{96}{\kilo\parsec}$ vertically.
The cavity has a radius of $r_{\rm b} = \SI{8}{\kilo\parsec}$ with
a density of $\rho_{\rm b} = \SI{2.5e-26}{\gram\per\cubic\cm}$ and temperature of
$T_{\rm b} = \SI{4e6}{\kelvin}$.
The surrounding medium at $z = 0$ has a density of $\rho_0 = \SI{1e-25}{\gram\per\cubic\cm}$
and temperature of $T_0 = \SI{1e6}{\kelvin}$.
On the system acts the gravitational acceleration of $g = \SI{3.0985e-7}{\cm\per\square\second}$.
Our simulations then run between $\SI{200}{\mega\year}$ and $\SI{250}{\mega\year}$.
The amplitude of the magnetic field is between $B_0 = \SI{2.5e-6}{\gauss}$ % and $B_0 = \SI{1e-5}{\gauss}$.
and $B_0 = \SI{6.39e-4}{\gauss}$
For the viscosity we obtain $\nu = \SI{3.0172e27}{\square\cm\per\second}$ and resistivity
$\eta = \SI{9.0516e26}{\square\cm\per\second}$.

\begin{table}
\centering
\begin{tabular}{l|l}
one code unit of & physical unit \\ \hline
length & $\SI{10}{\kilo\parsec}$ \\
time & $\SI{10}{\mega\year}$ \\
density & $\SI{e-25}{\gram\per\cubic\cm}$ \\
temperature & $\SI{e6}{\kelvin}$ \\
magnetic field & $\SI{e-4}{\gauss}$
\end{tabular}
\caption{Code unit conversion table.}
\label{tab: unit conversions}
\end{table}

\section{Effect of an Internal Helical Magnetic Field}
\label{sec: internal}

\subsection{Non-magnetic test case}
\begin{figure}[t!]\begin{center}
\includegraphics[width=0.95\columnwidth]{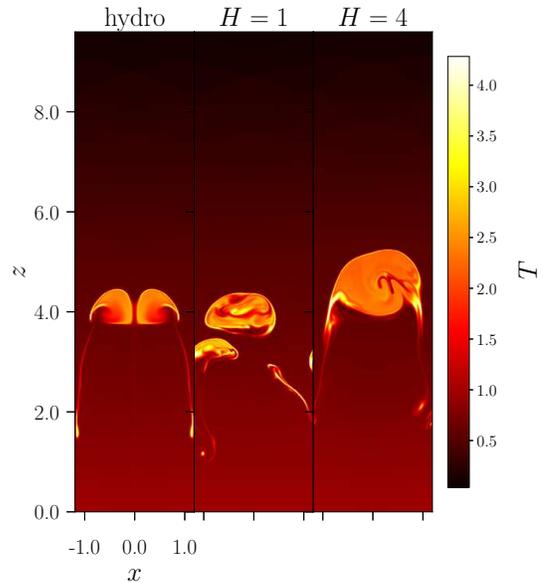}
\end{center}
\caption[]{
Slices through the simulation domain showing the temperature distribution at final times ($t = 20$)
for the purely hydrodynamical case (left), and for the magnetic scenario with ABC field in the weak helicity case (center) and strong helicity case (right). They corresponds to models hydro, hel\_l and hel\_h in table \Tab{tab: simulations}.
}
\label{fig: TT_t_20_nu_1e-3_sharp_highRes}
\end{figure}

The first case we take into account is the the purely hydrodynamical one, 
to which we will compare the simulations including magnetic fields.
As the cavity rises the Kelvin--Helmholtz instability acts on its interface with the
surrounding medium and
eventually affects the entire cavity by non-linear Kelvin--Helmholtz instability induced turbulent
mixing, as illustrated in the left panels of \Fig{fig: TT_t_20_nu_1e-3_sharp_highRes},
\Fig{fig: emission_t_20_nu_1e-3_sharp_highRes}
% \Fig{fig: TT_t_8.0_nu_2e-4_spheromak}, \Fig{fig: emission_t_8_nu_2e-4_spheromak}
and \Fig{fig: lnTT_t11}.

 \subsection{ABC Configuration}
We then perform two different simulations including an ABC helical magnetic field (see Section \ref{Sec:ABC}) inside the cavity.
There it becomes evident that magnetic helicity contributes to keep the cavity in a significantly
more coherent state and prevents its disruption, provided a minimum amount
of helicity is present in the field.
This is evident from the central and left panels of 
\Fig{fig: TT_t_20_nu_1e-3_sharp_highRes}, showing two helical cases: the amount of magnetic helicity 
in the simulation shown in the right panel is 4 times larger than that in the central panel.

\begin{figure}[t!]\begin{center}
\includegraphics[width=0.95\columnwidth]{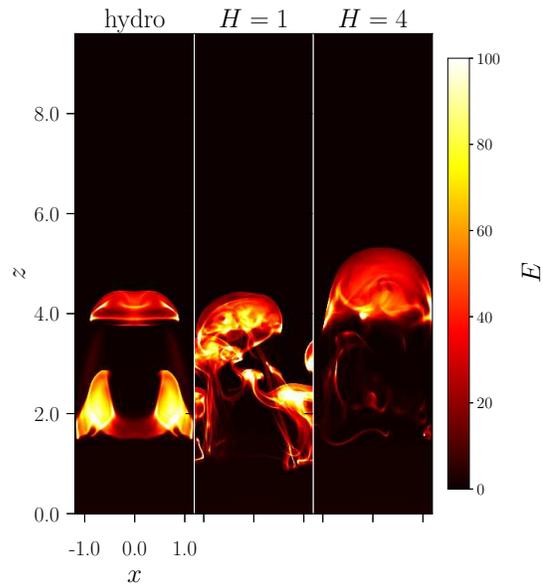}
\end{center}
\caption[]{
Emission measure at final times ($t = 20$)
for the purely hydrodynamical case (left), and for the magnetic scenario with ABC field in the weak helicity case (center) and strong helicity case (right).  They correspond to models hydro, hel\_l and hel\_h in table \Tab{tab: simulations}.}
\label{fig: emission_t_20_nu_1e-3_sharp_highRes}
\end{figure}

In order to make it easier to compare our results with observations we create artificial
emission measures, assuming an optically thin medium.
We place the observer along the $y$ axis at an infinite distance.
The emission measure $E$ is then simply the line integral of the temperature to the fourth power
\EQ\label{eq: emission}
E(x, z) = \int T^4\ \dd y.
\EN
Similar to the slices plotted in \Fig{fig: TT_t_20_nu_1e-3_sharp_highRes}, 
we observe from the emission measures
(\Fig{fig: emission_t_20_nu_1e-3_sharp_highRes}) an increased stability
when a helical magnetic field is present.

\begin{figure*}[t!]\begin{center}
\includegraphics[width=1.9\columnwidth]{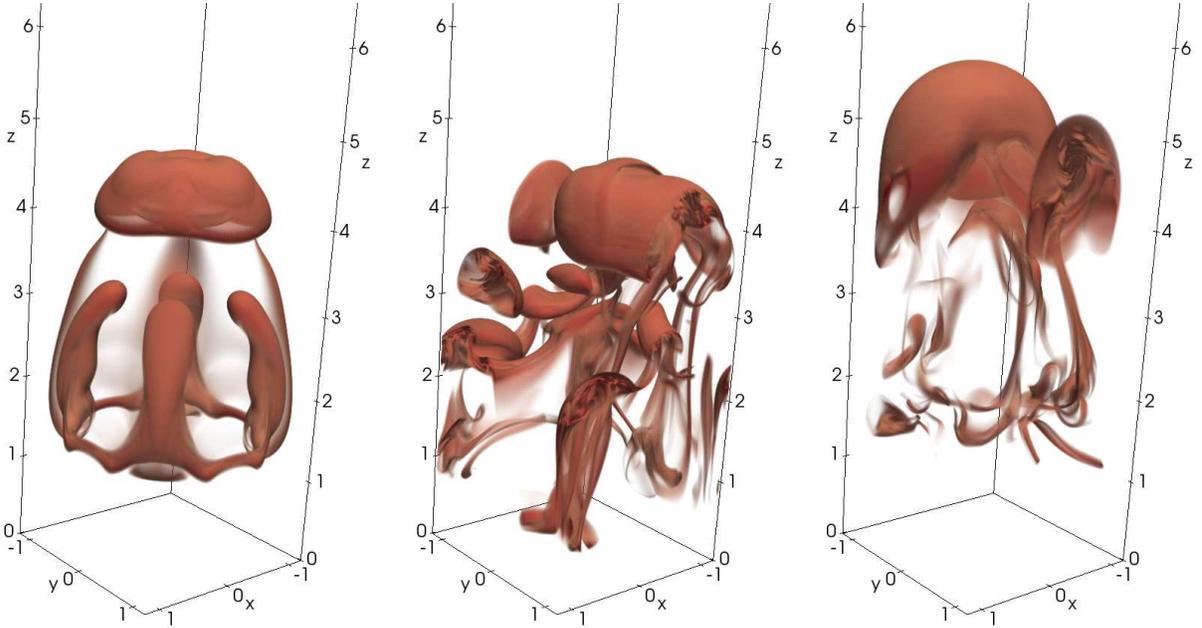}
\end{center}
\caption[]{
Volume rendering of the temperature of the magnetic cavity for the hydrodynamical case (left),
low magnetic helicity (center) and high magnetic helicity (right) at $t = 20$ for the ABC configuration.
It is evident that magnetic helicity plays a role in determining the topology of the cavity.
In particular, the symmetry seen in the hydrodynamical case is broken when magnetic helicity is non-zero.
Note that the four-fingers asymmetry in the hydrodynamical case is due to the geometry and finite size of the computational domain.
}
\label{fig: lnTT_t11}
\end{figure*}

We can also observe this behavior in the volume rendering of the temperature (\Fig{fig: lnTT_t11}).
While the purely hydrodynamical case (left panel) and the low-helicity case (center) disintegrate after two bubble
diameter crossings, the strongly helical case (right) remains largely intact.
Furthermore, in the purely hydrodynamical case, the temperature remains symmetric about the
central axis of the domain, as expected from symmetry properties of our configuration.
This is in agreement with simulations of \cite{Ruszkowski-Ensslin-2007-378-662-MNRAS} as well as
\cite{Dong-Stone-2009-704-1309-ApJ}, who both produced perfectly symmetric configurations in the hydrodynamical case.
Conversely, the disruption of the cavity in the low-helicity case develops a very asymmetric, chaotic structure.
Also in the high-helicity case, although the disruption is only marginal, we can still observe an
asymmetric evolution of the cavity.

\begin{figure}[t!]\begin{center}
\includegraphics[width=0.95\columnwidth]{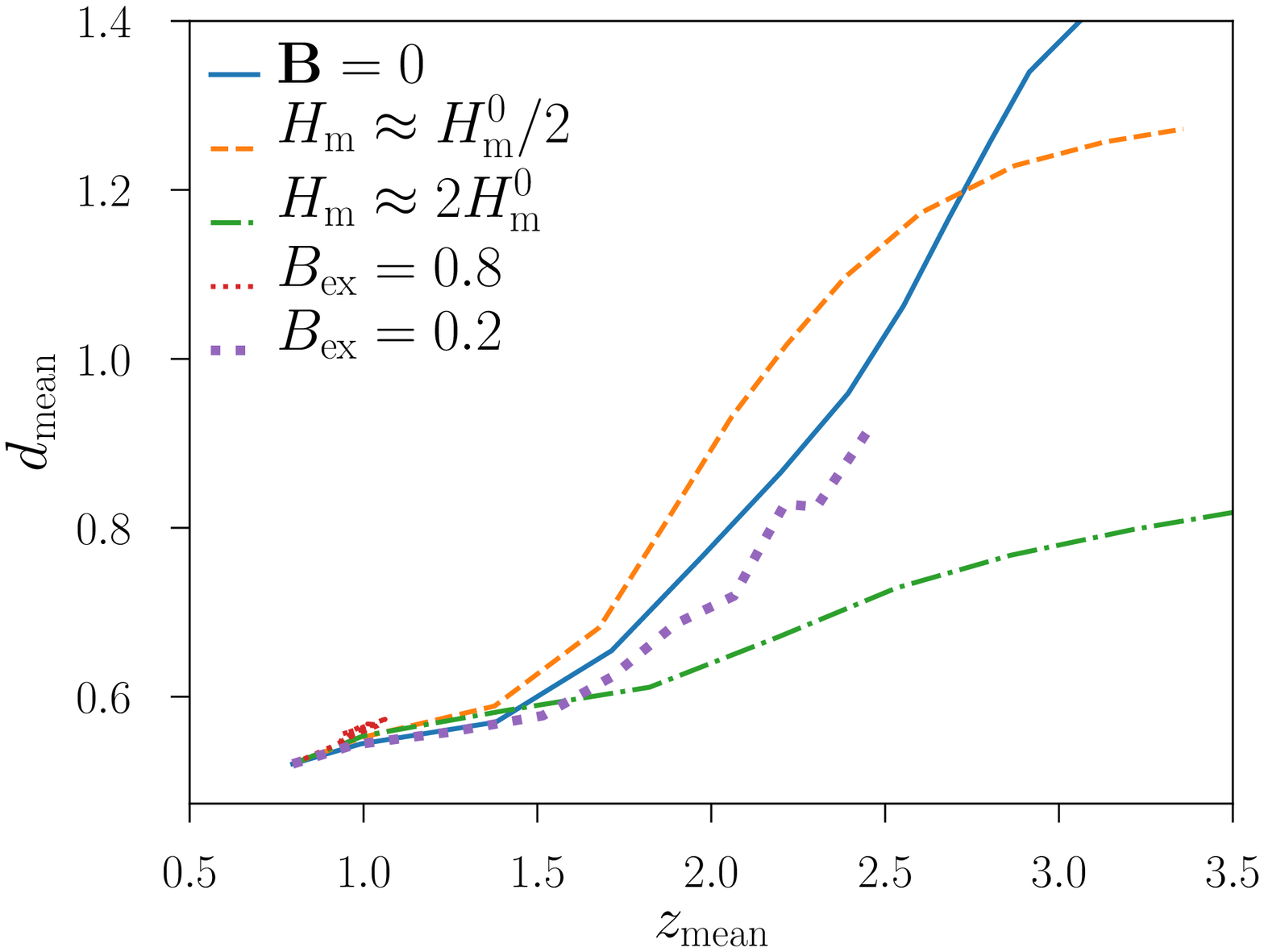} \\
\includegraphics[width=0.95\columnwidth]{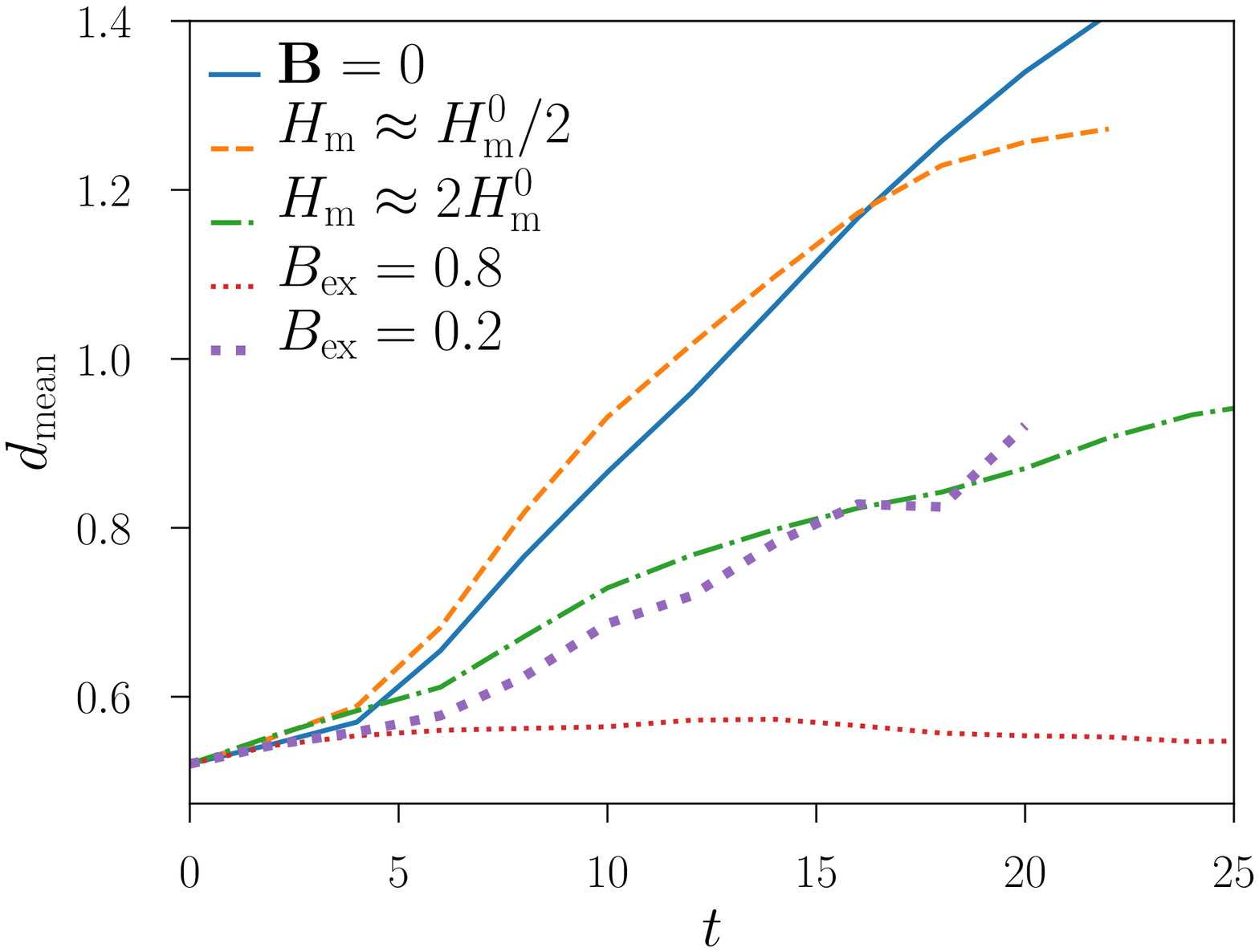}
\end{center}
\caption[]{
Coherence measure $d_{\rm mean}$ of the cavities in dependence of the mean height (upper panel)
and time (lower panel) for the hydrodynamical, weak and strong-helicity ABC cases and weak and strong external field cases.
We clearly observe that a large amount of magnetic helicity increases the cavities' stability.
A comparable effect can only be obtained with a substantially stronger external magnetic field, but such a field
inhibits the ascent of the bubble in the intergalactic medium.
The case of very strong external parallel field keeps the bubble very stable, but it
completely obstructs its ascent, hence keeping it confined in the lowest part of the domain.
Note that the high helicity and strong external field simulations continue well beyond the plotted time limit.
}
\label{fig: coherence_TT_dmean}
\end{figure}

To test the stability of the cavities we measure their coherence.
In order to do so, we start by defining the space filled by the cavity as the loci for which $\log_{10}(T) > 1.5$.
We choose this threshold because the surrounding cold medium has a significantly lower temperature
and in the simulated times we do not observe a high enough temperature diffusion or conduction
that would reduce the cavity temperature below this value.
We then measure the mean distance $d_{\rm mean}$ of all the points in the cavity
\EQ \label{eq: d_mean}
d_{\rm mean} = \langle\left|\rr_{\rm cavity} - \rr_{\rm CM}\right|\rangle,
\EN
where $\rr_{\rm CM}$ is the position vector to the center of mass defined as
\EQ
\rr_{\rm CM} = \frac{\int_{\log_{10}(T) > 1.5} T\rr \dd V}{\int_{\log_{10}(T) > 1.5} T \dd V},
\EN
where $\rr_{\rm cavity}$ is the position vector to a point within the cavity, i.e.\ $\log_{10}(T) > 1.5$,
and we take the average over the entire domain.
We can therefore study the evolution of $d_{\rm mean}$ with time.
Since some of the bubbles rise at different speeds, due to the different
magnetohydrodynamic parameters of different models, we can also study the behavior of $d_{\rm mean}$ as function
of the height reached by the bubble, i.e. by its center of mass.
For that we also compute the mean height of the bubbles as
\EQ \label{eq: z_mean}
z_{\rm mean} = \langle\left|z_{\rm cavity} - z_{\rm CM}\right|\rangle,
\EN
where we use the loci for which $\log_{10}(T) > 1.5$, similar as we do for $d_{\rm mean}$.

\Fig{fig: coherence_TT_dmean} (upper panel) depicts $d_{\rm mean}$ as function of $z_{\rm mean}$ in our simulations.
We observe that without a magnetic field the
cavity is dispersed without reaching far away from its starting position.
An internal magnetic field with a low helicity content (model hel{\_}l in \Tab{tab: simulations})
does not improve significantly the stability for almost the whole simulation, but it has a stabilizing effect towards the end.
However, an internal field with higher magnetic helicity (model hel\_h in \Tab{tab: simulations}),
and the same magnetic energy has some clear positive effect on the bubble's stability.
In this latter case we observe a relatively stable rise throughout.
\Fig{fig: coherence_TT_dmean} (lower panel) depicts instead the time evolution of $d_{\rm mean}$.
Here too it is evident how the strong-helical case is the most stable,
whilst the low helical one only marginally stabilizes the cavity.
From \Fig{fig: coherence_TT_dmean} we
can see that high helicity stabilizes the cavity for at least \SI{250}{\mega\year}, whilst the hydrodynamical
case and the low-helicity case disrupt after about \SI{100}{\mega\year}.

% \begin{figure}[t!]\begin{center}
% \includegraphics[width=0.95\columnwidth]{coherence_TT_dmean_t}
% \end{center}
% \caption[]{
% Coherence measure $d_{\rm mean}$ of the cavities in dependence of time for the hydrodynamical,
% weak and strong helicity case and weak and strong external field cases.
% The weak helicity case differs from the the hydro case in the last part of the simulation.
% The strong helicity case shows a clear increase of the bubble stability, similar to that of an external parallel field. 
% The case of very strong external parallel field keeps the bubble very stable, but it
% completely obstructs its ascent, hence keeping it confined in the lowest part of the domain.
% Note that the high helicity and strong external field simulations continue well beyond the plotted time limit.
% }
% \label{fig: coherence_TT_dmean_t}
% \end{figure}

\subsection{Behavior at Higher Reynolds Number}

\begin{figure}[t!]\begin{center}
\includegraphics[width=0.95\columnwidth]{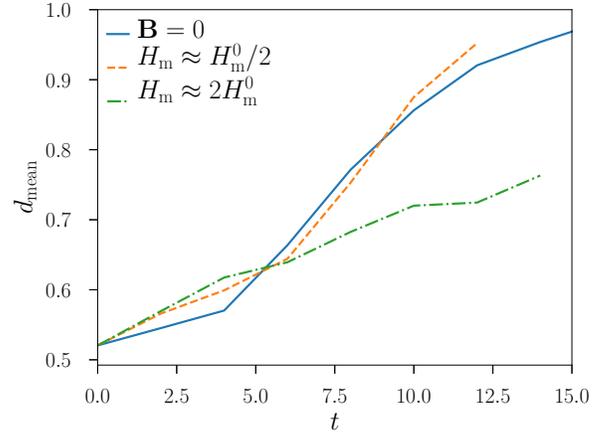}
\end{center}
\caption[]{
Coherence measure $d_{\rm mean}$ of the cavities for the high fluid Reynolds number cases
in dependence of time for the hydrodynamical,
weak and strong helicity ABC case.
The weak helicity case differs from the the hydro case in the last part of the simulation.
The strong helicity case shows a clear increase of the bubble stability,
confirming our low Reynolds number results.
}
\label{fig: coherence_TT_dmean_t_lowNu}
\end{figure}

We test the effect of a higher fluid Reynolds number on our results by performing simulations
with a lower viscosity, $\nu = 2\times10^{-4}$ for the hydrodynamical and the helical cases
(models hydro2, hel\_l2, and hel\_h2 in table \Tab{tab: simulations}).
There we observe a clearer onset of the Kelvin--Helmholtz instability.
However, the coherence measure does not change significantly for any of the models compared to the
low Reynolds number case, and the relative behavior between the hydrodynamical
case and the two cases with internal helical magnetic fields resembles that obtained with
higher viscosity (\Fig{fig: coherence_TT_dmean_t_lowNu}).
Therefore, this confirms our results in the lower Reynolds number regime.

\subsection{Spheromak Configuration}

In order to eliminate effects from the particular geometry of the magnetic field, we test the
stabilizing properties of the spheromak magnetic field configuration from equation \eqref{eq: bb spheromak}.
To obtain a fair comparison with the ABC field case, we want to analyze a scenario with the same amount of
magnetic energy and magnetic helicity.
To do so, we choose appropriate values for the field amplitude and parameter $\tau$,
which in this case plays the same role of the parameter $k$ for the ABC field.
Since we have two degrees of freedom we can match both, the magnetic energy and the helicity contents,
so that our low (high) helicity spheromak configuration has almost the same magnetic energy and helicity as the
low (high) helicity ABC configuration (see sph{\_}l and sph{\_}h in \Tab{tab: simulations}).

From the parameters in \Tab{tab: simulations} we can see that our spheromak simulations belong to a high
$\Rey$ and $\Rm$ scenario.
Since the parameters $\nu$ and $\eta$ are the same used in the high Reynolds number cases for the ABC
field we can deduce that higher values of $u_{\rm max}$ are reached.
Due to these localized high velocities of the gas that we attribute to local currents, we are able to
run the high-helicity simulation only to a simulation time of ca.\ $9$.
Nevertheless, this gives us enough data to confirm the results obtained through the ABC configuration.
That is, a high magnetic helicity internal magnetic field can stabilize the bubbles.

\begin{figure}[t!]\begin{center}
\includegraphics[width=0.95\columnwidth]{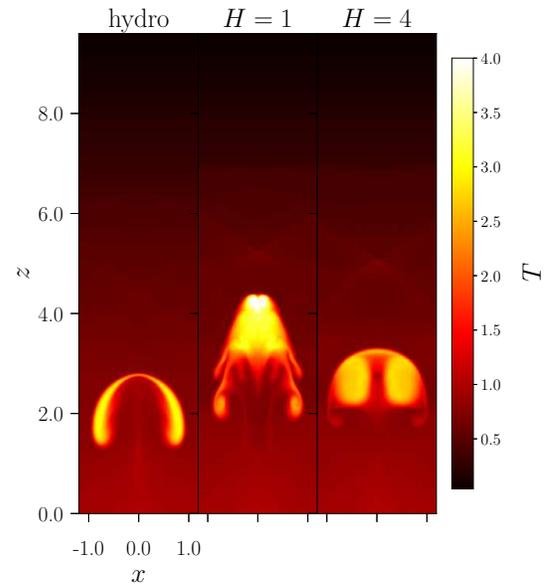}
\end{center}
\caption[]{ 
Slices through the simulation domain showing the temperature distribution at times ($t = 8$)
for the purely hydrodynamical case (left), weak helicity spheromak case (center)
and strong helicity spheromak case (right)
(models hydro, sph\_l and sph\_h in table \Tab{tab: simulations}).
}
\label{fig: TT_t_8.0_nu_2e-4_spheromak}
\end{figure}

\begin{figure}[t!]\begin{center}
\includegraphics[width=0.95\columnwidth]{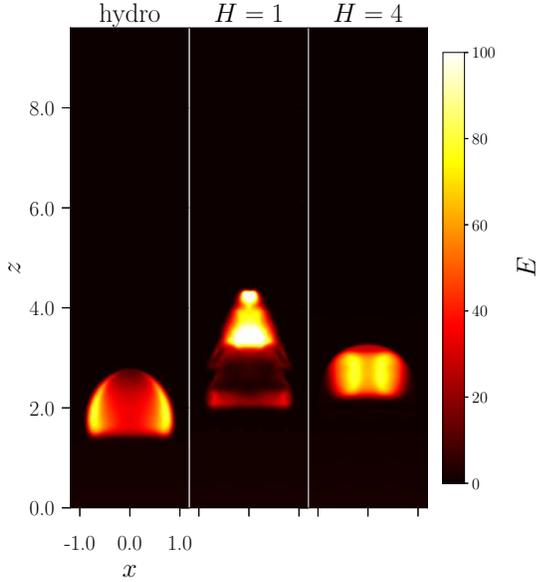}
\end{center}
\caption[]{
Emission measure at times ($t = 8$)
for the purely hydrodynamical case (left), weak helicity spheromak case (center)
and strong helicity spheromak case (right).
They corresponds to models hydro, sph\_l and sph\_h in table \Tab{tab: simulations}.}
\label{fig: emission_t_8_nu_2e-4_spheromak}
\end{figure}

\begin{figure}[t!]\begin{center}
\includegraphics[width=0.95\columnwidth]{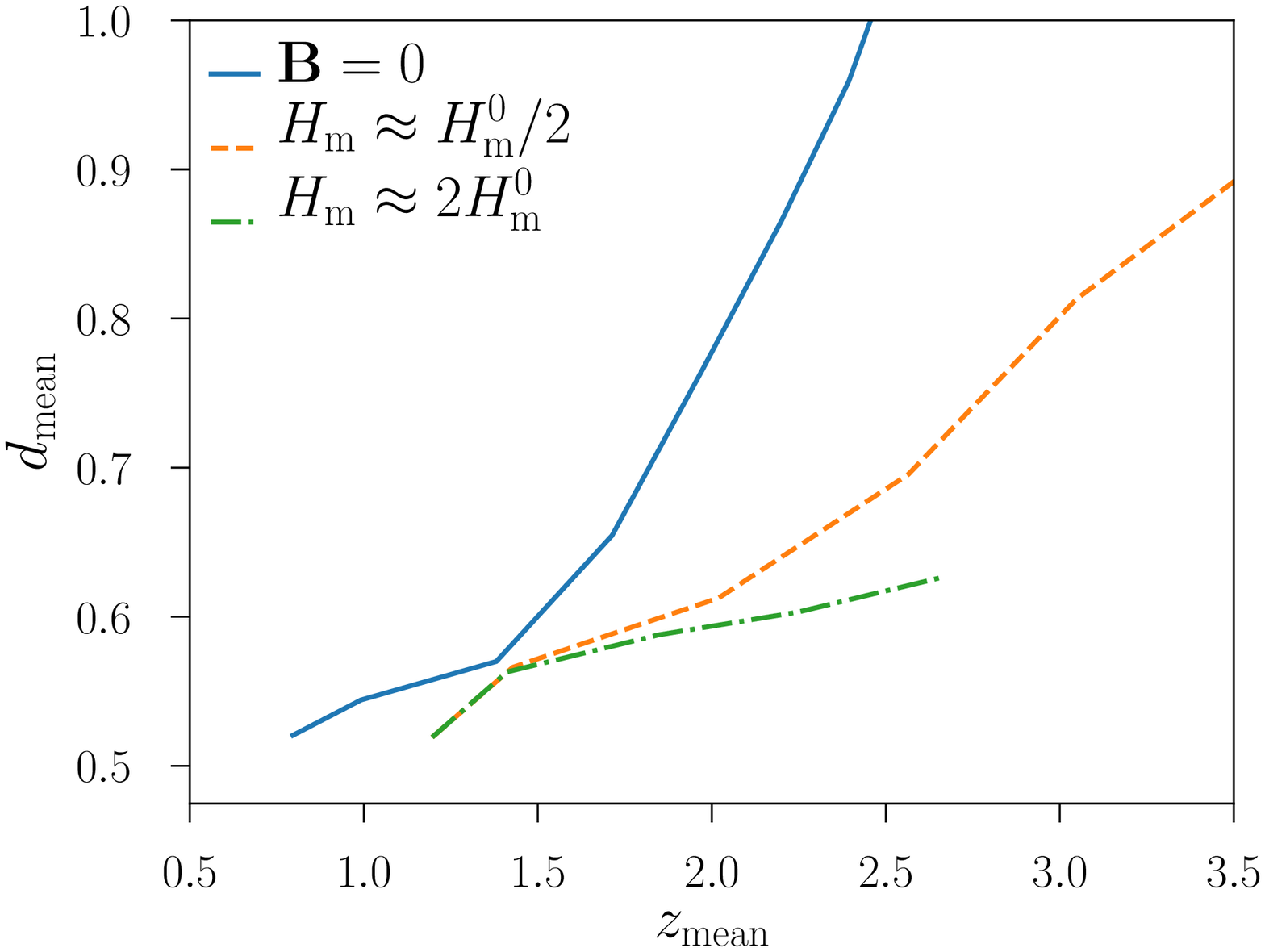} \\
\includegraphics[width=0.95\columnwidth]{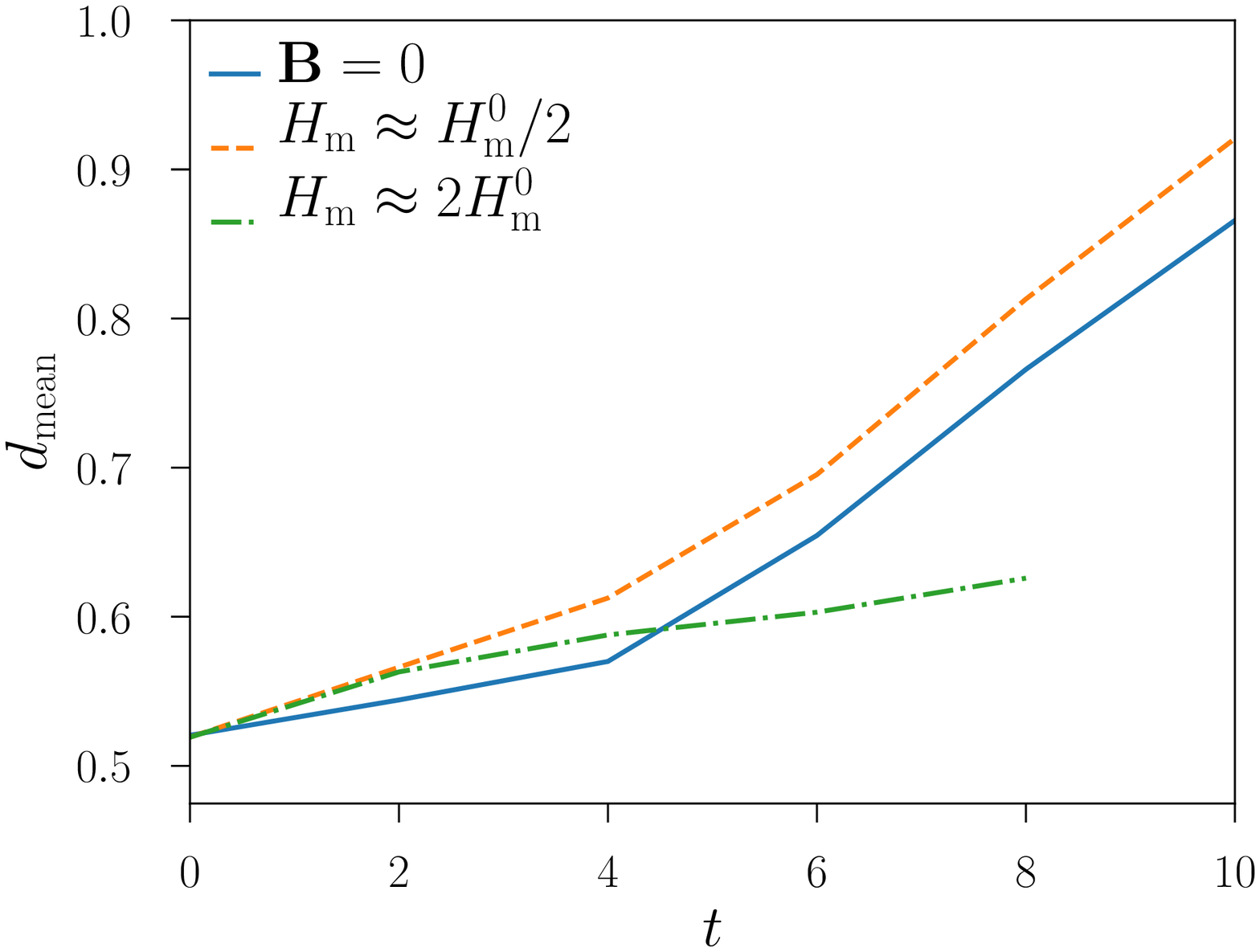}
\end{center}
\caption[]{
Coherence measure $d_{\rm mean}$ of the cavities for the high fluid Reynolds number
and hydrodynamical case, low and high magnetic helicity spheromak cases in dependence of mean height
(upper panel) and time (lower panel).
Although the high helicity simulation stops at time $9$ we can still clearly see that
an internal helical magnetic field stabilizes the bubbles and prevents their timely disruption.
A small amount of helicity is clearly not sufficient.
}
\label{fig: coherence_TT_dmean_sph}
\end{figure}

We observe this stabilizing effect from the slices of the temperature (\Fig{fig: TT_t_8.0_nu_2e-4_spheromak})
and the emission measure (\Fig{fig: emission_t_8_nu_2e-4_spheromak}).
The high helicity spheromak case clearly stays  more stable.
From the coherence calculations depicted in \Fig{fig: coherence_TT_dmean_sph}, where we compare the hydrodynamical case
with the low and high helicity spheromak cases, this stabilizing effect is evident as well,
which is clearly visible in the high-helicity configuration.
Conversely, the low-helicity configuration is characterized by a coherence similar to the hydrodynamic case.
However, from the lower panel of \Fig{fig: coherence_TT_dmean_sph} we notice how in the low-helicity case
the bubble can reach higher parts of the domain. 

\section{Effect of an External Parallel Magnetic Field}
\label{sec: external}

From \cite{Chandrasekhar1961hhs} and \cite{Sharma-Srivastava-1968-21-917-AusJPhys}
we know that the Kelvin-Helmholtz instability is suppressed by a magnetic field
that is aligned (parallel) to the velocity.
Comparing equation (13) in chapter IV from \cite{Chandrasekhar1961hhs} with our induction
equation \eqref{eq: induction} we observe that we can identify our $\BB$ with $\HH$ and
the magnetic resistivity $\eta$ has the same definition and the permeability $\mu = 1$ in our formulation.

With that we can write the equation (205) in chapter XI from \cite{Chandrasekhar1961hhs} as
\EQ \label{eq: suppressed KH}
B^2 \ge 2\pi (u_1 - u_2)^2(\rho_1\rho_2)/(\rho_1 + \rho_2),
\EN
which gives us a criterion for the Kelvin-Helmholtz instability to be entirely
suppressed for a two-layer system with densities $\rho_1$ and $\rho_2$ and velocities
$u_1$ and $u_2$.
For our hydrodynamic simulations we observe a velocity difference of ca.\ $0.5$,
while our densities are $1$ (surrounding medium) and $0.25$ (hot cavity).
With that we estimate the magnetic field strength parallel to the velocity
to be ca.\ $0.56$ for suppressing the Kelvin-Helmholtz instability.

Here we present the evolution of such a cavity with $\BB = 0.8 \ee_z$
and one with $\BB = 0.2 \ee_z$.
That is, one will exhibit Kelvin-Helmholtz suppression at all scales, while the
other will not.
For a field that is strong enough for suppression we are in a situation
with little buoyancy, that is $z_{\rm mean}$ remains approximately constant in time
(\Fig{fig: coherence_TT_dmean}).
It is worth noticing that both these cases with a magnetic field along $\ee_z$ have a total magnetic
energy between one and two order of magnitude higher than the helical cases.

\section{Discussion and conclusions}
\label{sec: conclusion}

In this work we have examined the possibility that a helical magnetic field may play a key role in the stability
of extragalactic bubbles, similar to the Fermi bubbles observed raising from the midplane of our
Galaxy in the intergalactic medium.
This hypothesis appears justified, because these bubbles are thought to be inflated by AGN or
from jets coming from the galactic center.
Such jets have been observed to be characterized by helical magnetic fields, which is a consequence of the
rotation of strong magnetic field from their sources.
Since magnetic helicity is conserved in a high-conductivity medium, such as the intergalactic medium,
it is reasonable to think the bubbles raising in the intergalactic
medium to retain their helical magnetic field.
As we used parameters that can be compared to measurements from the intergalactic medium we can directly
draw an analogy between our simulations and observations.

For the purely hydrodynamical case we observe a longer stability (ca.\ \SI{80}{{\mega\year}})
than has been predicted for bubbles in the intergalactic medium, although with an increase
of ca.\ $50$\% in the coherence measure $d_{\rm mean}$.
While a parallel magnetic field is known to suppress the Kelvin-Helmholtz instability,
for the case of intergalactic magnetic cavities the field strength would need to be rather large.
Here we have shown that we can exploit the stability properties of magnetically helical
structures to keep these cavities from disrupting, with a much smaller magnetic energy content.

We used a general helical magnetic field in the form of the ABC flow
that fills a bubble raising through buoyancy in an otherwise stably stratified medium.
We quantify the disruption of the bubble by measuring the parameter $d_{\rm mean}$,
the mean distance of all the points contained in it.
We observed that this bubble is stabilized and does not develop a Kelvin-Helmholtz instability
in the interface with the surrounding medium if the magnetic field is sufficiently helical.
We estimated that a helical field with maximum strength of the order of $\SI{e-5}\gauss$ can
stabilize the bubble over a time scale of about \SI{250}{\mega\year}.
Here we see that during that time the high helicity case never exceeds a value of
$d_{\rm mean}$ twice its initial value.
Conversely, a less helical magnetic field, with a total magnetic helicity 4 times lower, cannot keep
the bubble stable and this case exhibits a disruption similar to the non-magnetic case.
In the low-helicity and hydrodynamic cases
$d_{\rm mean}$ increases by more than twice its initial value.

To verify that our results do not depend on a specific initial geometry, we performed additional
simulations using a different initial magnetic field filling the bubble.
We used a spheromak field with the same initial content of magnetic energy and helicities.
By evaluating the coherence measure of the bubble we showed how the overall evolution of the
system is similar to the ABC cases, hence showing the important role played by magnetic helicity. 

Therefore, based on the results here presented, we propose that an internal helical magnetic field is a viable explanation of
intergalactic bubble stability, requiring only relatively low magnetic energies compared to the
external magnetic field hypothesis.

\section*{Acknowledgments}

The authors appreciate the support given by the HPC3 Europe program HPC-EUROPA3 (INFRAIA-2016-1-730897).
SC acknowledges financial support from the UK's STFC (grant number ST/K000993).
For the plots we made use of the Matplotlib library for Python \citep{Hunter:2007}
and BlenDaViz\footnote{github.com/SimonCan/BlenDaViz}.
The authors thank Department of Physics of the University of Crete for the hospitality
and Evangelia Ntormousi, Alexander Russell, David I. Pontin, Gunnar Hornig and Ross Palister for
constructive and critical discussions.
We thank the anonymous referee for the constructive critique that led to substantial improvements of this paper.

\bibliography{references}
\end{document}